\documentclass{acm_proc_article-sp}

\usepackage{xcolor}
\usepackage{graphicx} 
\usepackage{epstopdf}
\usepackage{subfigure}
\usepackage{listings}

\lstdefinelanguage{ACME}{
   morecomment = [l]{//}, 
   morecomment = [l]{///},
   morecomment = [s]{/*}{*/},
   morestring=[b]", 
   sensitive = false,
   keywords = {and,Attachments,boolean,Component,Connector,Dependencies,Detach,do,extends,extended,false,Family,Forall,from,in,Invariant,new,On,Port,Property,Remove,Role,System,true,to,Type,with}
}
\lstdefinelanguage{PIADL}{
   morecomment = [l]{--},
   morestring = [b]",
   sensitive = true,
   keywords ={architecture, component, connector, abstraction, is, behaviour, Behaviour, behaviour, Behaviour, and, or, xor, compose, choose, where, unifies, relays, type, port, assuming, protocol, connection, in, out, inout, via, true, false, send, receive, value, Any, any Union, unobservable, function, if, the, else, tuple, project, as, selecting, including, Sequence, Set, Bag, sequence, set, bag, Natural, done, parameter, replicate, value, location, locate, Integer, restrict, iterate, from, by, accumulate}
}
\lstdefinelanguage{C2SADL}{
   morecomment = [l]{//}, 
   morecomment = [l]{///},
   morecomment = [s]{/*}{*/},
   sensitive = true,
   keywords = {component, is, interface, top\_domain, in, out, bottom\_domain, parameters, null, behaviour, context, end, architecture, connectors, component\_instances, instantiates, topology, connector, top\_ports, bottom\_ports, Unweld, Weld, AddComponent, AddConnector, RemoveComponent, RemoveConnector, sutype}
}
\lstdefinelanguage{Wright}{
   morecomment = [s]{[}{]},
   sensitive = true,
   keywords = {and, or, Style, EndStyle, Component, Connector, Port, Role, Computation, Glue, Constraints, Configuror, EndConfiguror, new, attach, detach, remove, where, to, Interface, Type, Instances, Attachments, Configuration, End}
}
\begin{document}

\conferenceinfo{CAL'2012}{6\`eme Conf\'erence francophone sur les architectures logicielles, Montpellier, France}
\CopyrightYear{2012}

\title{Issues of Architectural Description Languages for Handling Dynamic Reconfiguration}

\numberofauthors{4}
\author{
  \alignauthor Leonardo A. Minora\\
    \affaddr{Federal Institute of Education, Science and Technology of Rio Grande do Norte}\\
    \affaddr{Natal (RN), Brazil}\\
    \email{leonardo.minora@ifrn.edu.br}
  \alignauthor J\'er\'emy Buisson\\
    \affaddr{UEB / \'Ecoles de St-Cyr Co\"etquidan / Universit\'e de Bretagne Sud}\\
    \affaddr{Guer, France}\\
    \email{jeremy.buisson@st-cyr.terre-net.defense.gouv.fr}
\and
  \alignauthor Flavio Oquendo\\
    \affaddr{UEB  / Universit\'e de Bretagne Sud}\\
    \affaddr{Vannes, France}\\
    \email{flavio.oquendo@univ-ubs.fr}
  \alignauthor Thais V. Batista\\
    \affaddr{Federal University of Rio Grande do Norte}\\
    \affaddr{Natal (RN), Brazil}\\
    \email{thais@dimap.ufrn.br}
}

\maketitle

\begin{abstract}
Dynamic reconfiguration is the action of modifying a software system at runtime.
Several works have been using architectural specification as the basis for dynamic reconfi\-guration.
Indeed ADLs (architecture description languages) let architects describe the elements that could be reconfigured as well as the set of constraints to which the system must conform during reconfiguration. 
In this work, we investigate the ADL literature in order to illustrate how reconfiguration is supported in four well-known ADLs: $\pi$-ADL, ACME, C2SADL and Dynamic Wright. 
From this review, we conclude that none of these ADLs: 
(i) addresses the issue of consistently reconfiguring both instances and types; 
(ii) takes into account the behaviour of architectural elements during reconfiguration; and 
(iii) provides support for assessing reconfiguration, e.g., verifying the transition against properties.
\end{abstract}


\section{Introduction}\label{section:introdution}
The disciplined software engineering relies on software architectures to describe systems~\cite{Bass2003}. For modeling a software architecture, the academy proposed the architecture description language (ADL) and their toolsets~\cite{Garlan1997b, Medvidovic2000, Shaw1996}.
Also, the industry has used ADLs to develop systems.
Industry examples are
(i) $\pi$-ADL has been used to architect and refine federated knowledge management systems in Engineering Ingegneria Informatica - Italy~\cite{Oquendo2004a}; 
(ii) and AADL (Architecture Analysis \& Design Language) is a standard language for the Society of Automotive Engineers~\cite{SAE-AS5506}.
According to the state-of-the-art~\cite{Bass2003, Medvidovic2000, Mehta2000, Shaw1996}, the following concepts are relevant to describe software architectures: components and connectors (respectively computational and communication elements), architectural constraints, non functional properties, and behaviour.

Software systems evolve over their life time~\cite{Bennett2000a, Morrison2007, Oreizy1998a}.
Dynamic reconfiguration is when the evolution is performed at runtime with no service disruption. 
The dynamic reconfiguration can be handled by architectural concepts in an ADL.
However among many existing ADLs, only few allow modeling dynamic reconfiguration.
$\pi$-ADL~\cite{Oquendo2004}, ACME/Plastik~\cite{Batista2008a, Joolia2005}, C2SADL~\cite{Medvidovic1996, Oreizy1998}, DAOP-ADL~\cite{Fuentes2003}, Darwin~\cite{Magee1996}, Dynamic Wright~\cite{Allen1997, Allen1998a}, Rapide~\cite{Vera1999}, Weaves~\cite{Oreizy1999}, and xADL~\cite{Dashofy2002} are typical examples.
Nevertheless, there is no current consensus about how ADLs should address reconfiguration, e.g., what language constructs should be provided.

In this work, the goal is twofold: 
(i) to investigate the ADLs support for handling dynamic reconfiguration in the literature; and,
(ii) to illustrate how four well-known ADLs support dynamic reconfiguration: $\pi$-ADL, ACME, C2 SADL and Dynamic Wright.
We chose these four languages because they rely on different paradigms: the higher order typed $\pi$-calculus~\cite{Oquendo2004}, first order predicate logic~\cite{Garlan2000}, compo-nent- and event-based~\cite{Taylor1996a,Medvidovic1996}, and, graph grammars and communicating sequential processes (CSP)~\cite{Allen1998a,Allen1997}, respectively.
They also complement each other as $\pi$-ADL models the behavior of architectures, ACME focuses on the structure, C2SADL make the attention for components and their concurrent events, and Dynamic Wright supports the definition of structure and behavior.

This work is structured as follows.
Section~\ref{section:motivation} motivates this work thanks to the example of a TCP/IP stack system. 
Three scenarios illustrate three facets of dynamic reconfiguration.
Section~\ref{section:concepts} presents the basics concepts of software architecture and dynamic reconfiguration.
Section~\ref{section:related-works} details the related works about dynamic reconfiguration at the architecture level.
Section~\ref{section:comparison} compares the four above-mentioned ADLs in the light of our example of Section~\ref{section:motivation}.
Section~\ref{section:conclusion} concludes the paper with open issues that we note in the ADL support for dynamic reconfiguration.

\section{Motivation}\label{section:motivation}
Dynamic reconfiguration aims at modifying the software systems at runtime with no service disruption.
Critical systems usually want to benefit of a dynamic reconfiguration because any service disruption may have substantial consequences.
Examples of critical systems are stock market quotation systems, telecommunications systems, safety systems, and air traffic control systems.
However, there is no current consensus about how the software architecture should address dynamic reconfiguration at architecture level.
Does regular syntax for dynamic architectures allow the same reconfigurations as specific language constructs?
Should the state of components and connectors be taken into account at the architecture level?

\begin{figure}
\centering
\subfigure[TCP/IP Stack]{
   \epsfig{file=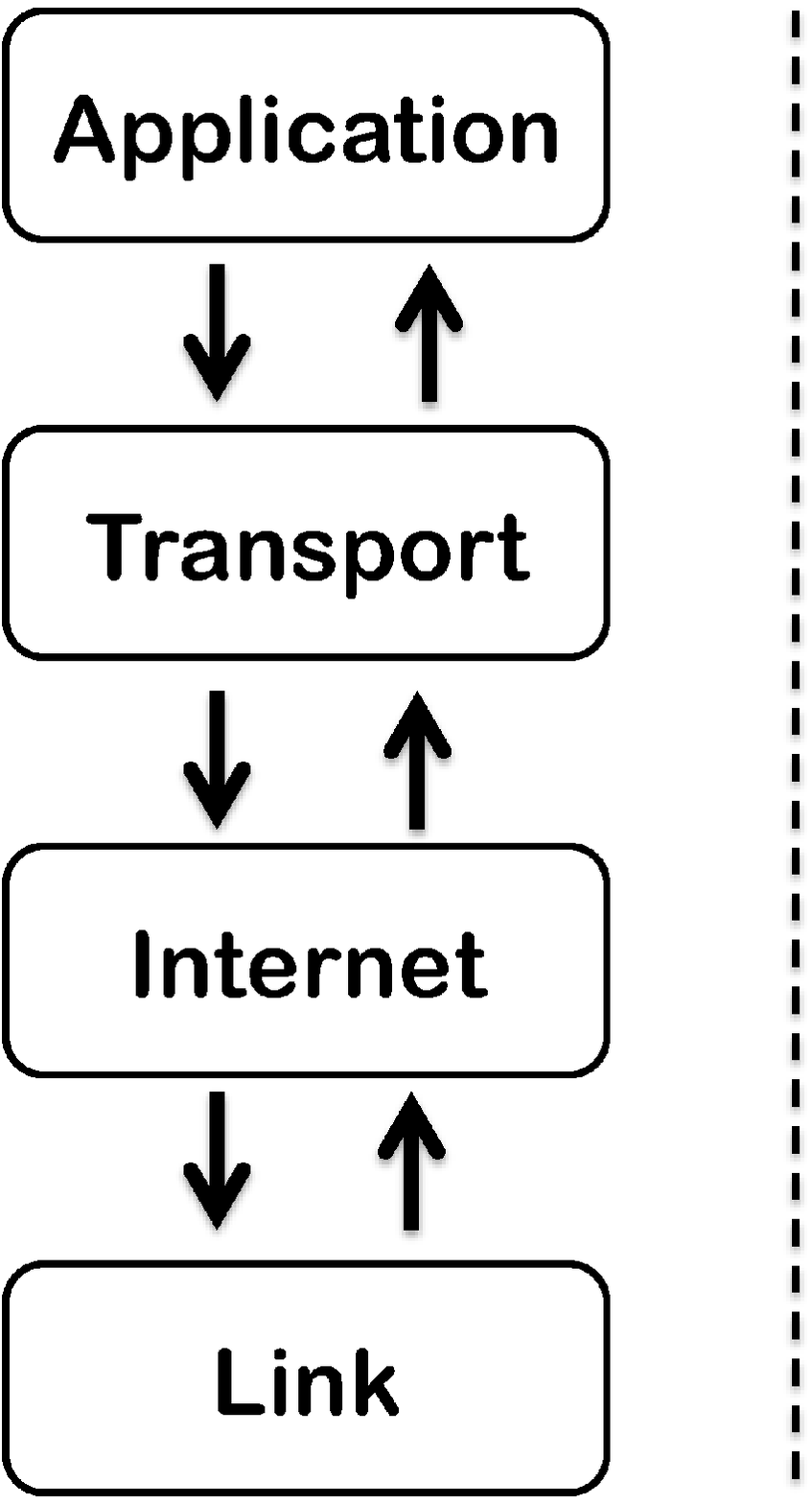, height=1.95in, width=1.05in}
   \label{fig:tcpip-stack}
}
\subfigure[TCP/IP Configuration]{
   \epsfig{file=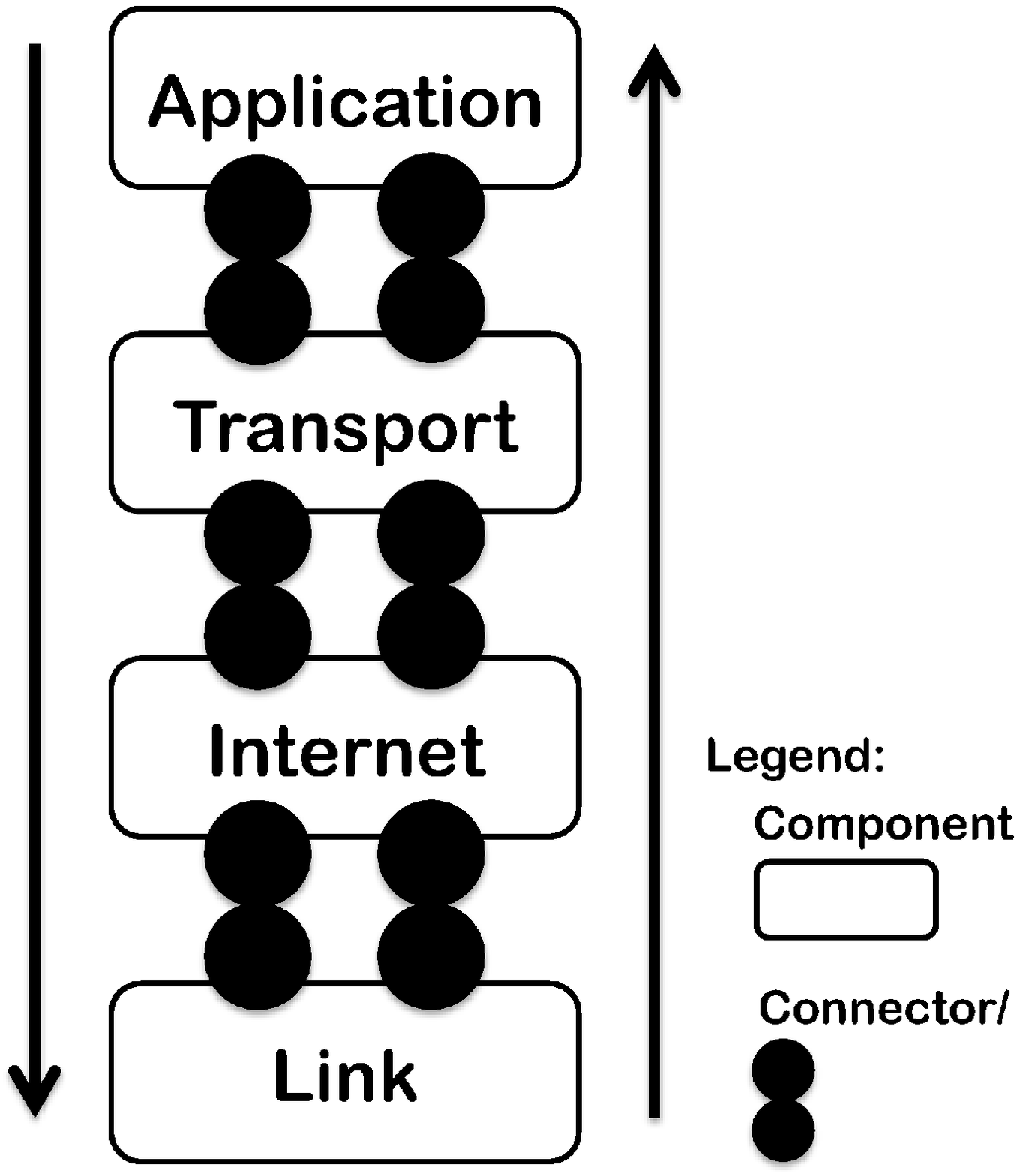, height=1.95in, width=1.74in}
   \label{fig:tcpip-system}
}
\caption{simplified version of the TCP/IP stack system.}
\end{figure}

To support this work, we rely on a simplified version of the TCP/IP stack (Fig.~\ref{fig:tcpip-stack}).
Each of the four layers (Application, Transport, Internet, and Link) is modeled as a component (Fig.~\ref{fig:tcpip-system}).

\begin{figure}
\centering
\subfigure[High bandwidth]{
   \epsfig{file=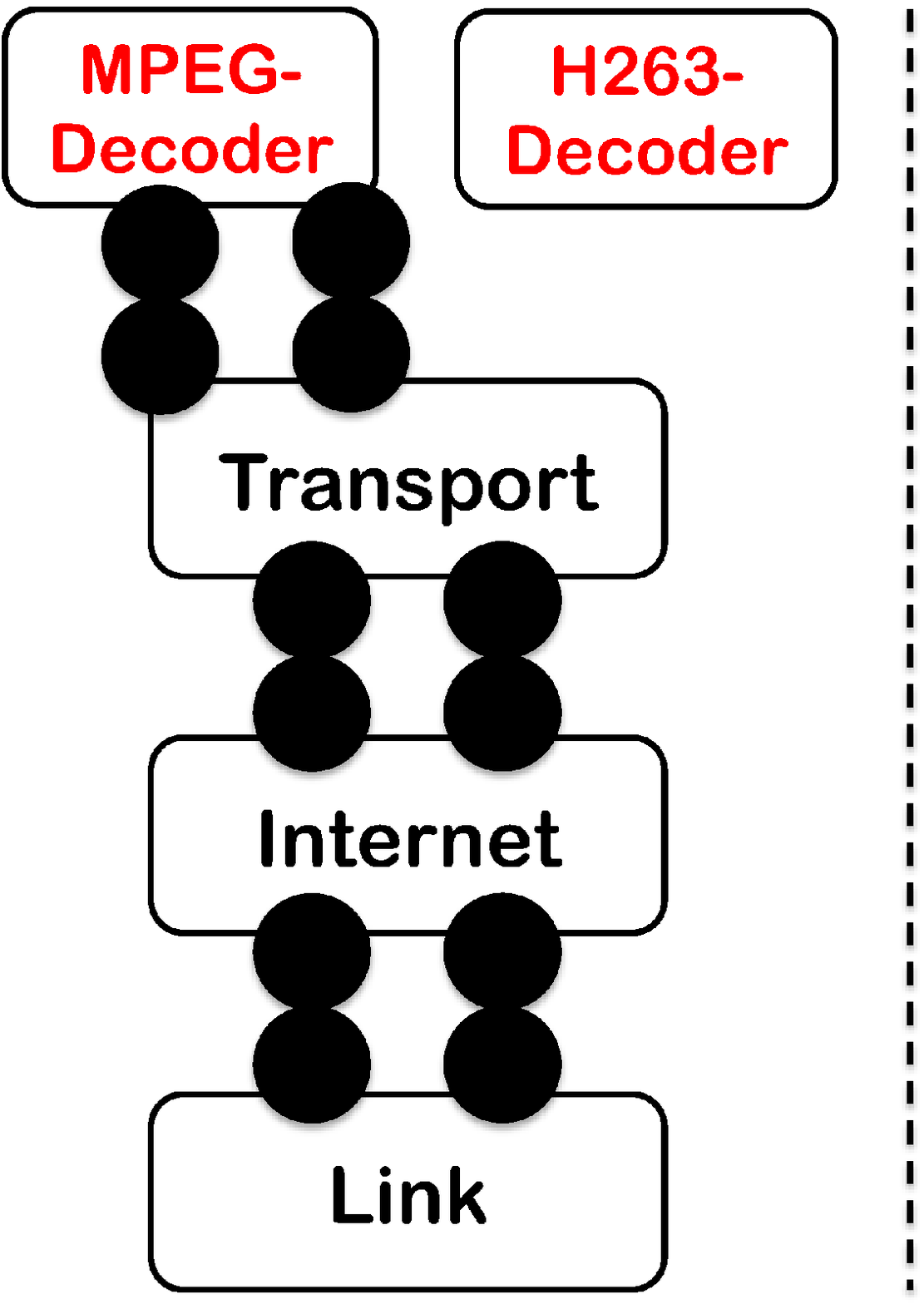, height=1.951in, width=1.38in}
   \label{fig:tcpip-mpeg}
}
\subfigure[Low bandwidth]{
   \epsfig{file=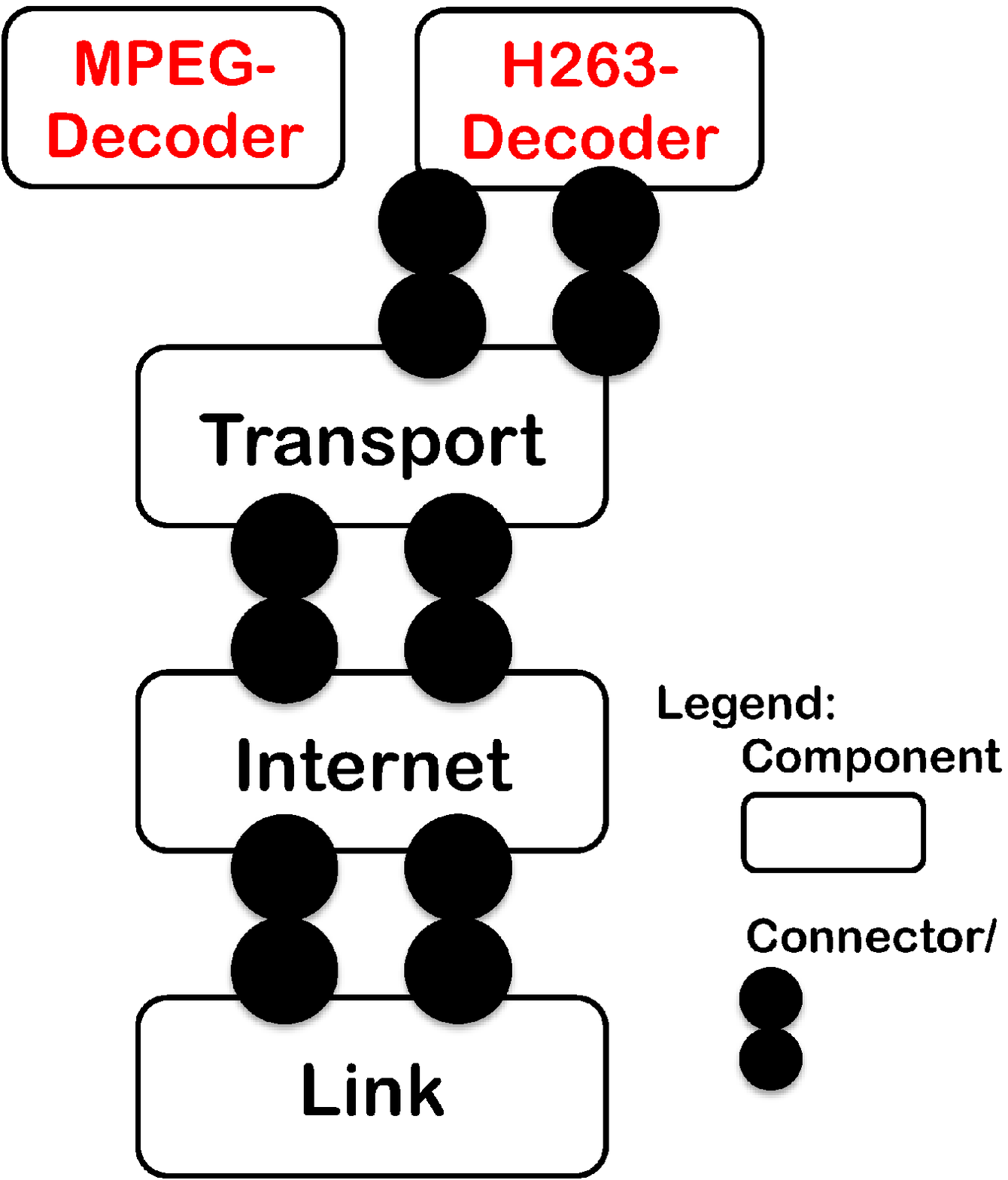, height=1.95in, width=1.70in}
   \label{fig:tcpip-h263}
}
\caption{scenario 1 to illustrate changing the component.}\label{fig:tcpip-scenario1}
\end{figure}

\begin{figure}
\centering
\subfigure[IPv4]{
   \epsfig{file=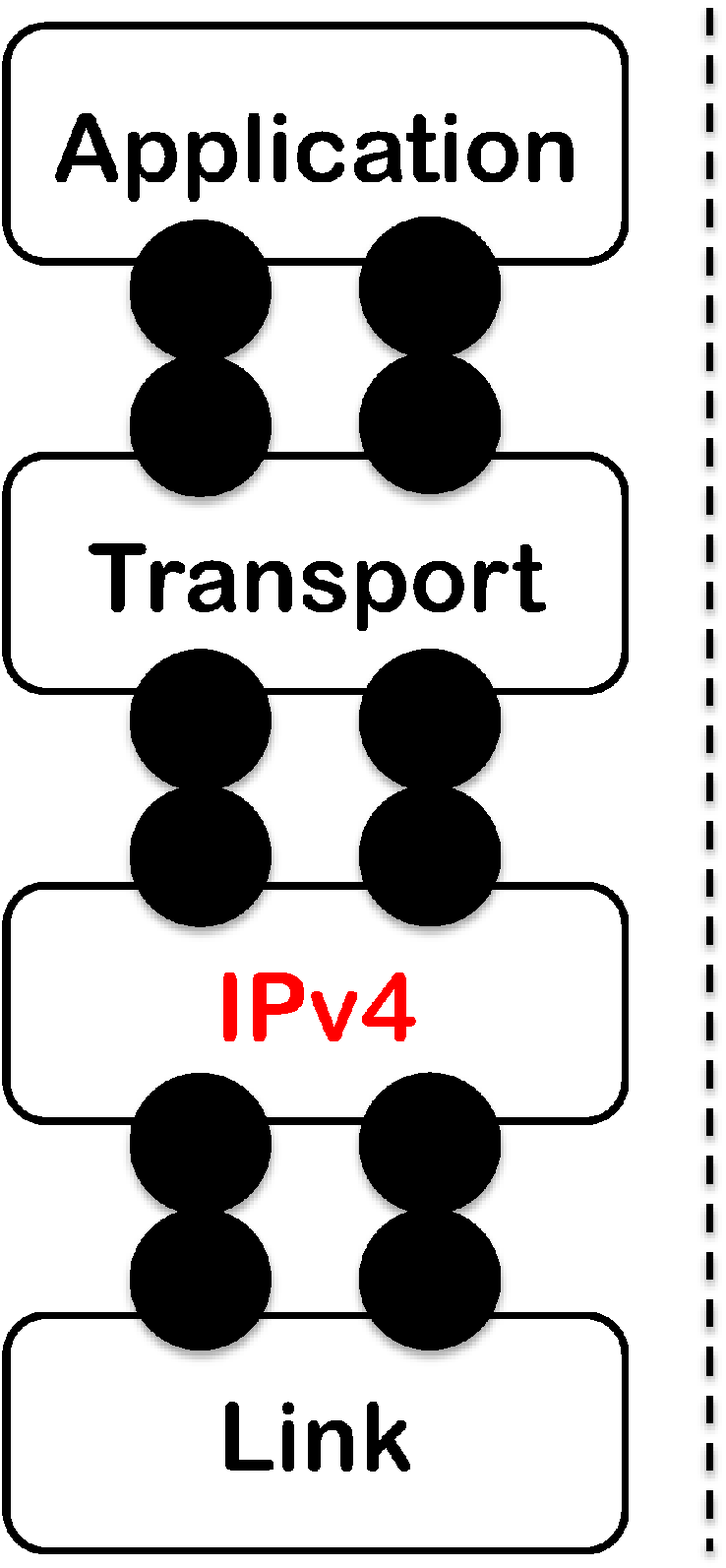, height=1.951in, width=0.90in}
   \label{fig:tcpip-ipv4}
}
\subfigure[IPv6 replaces IPv4]{
   \epsfig{file=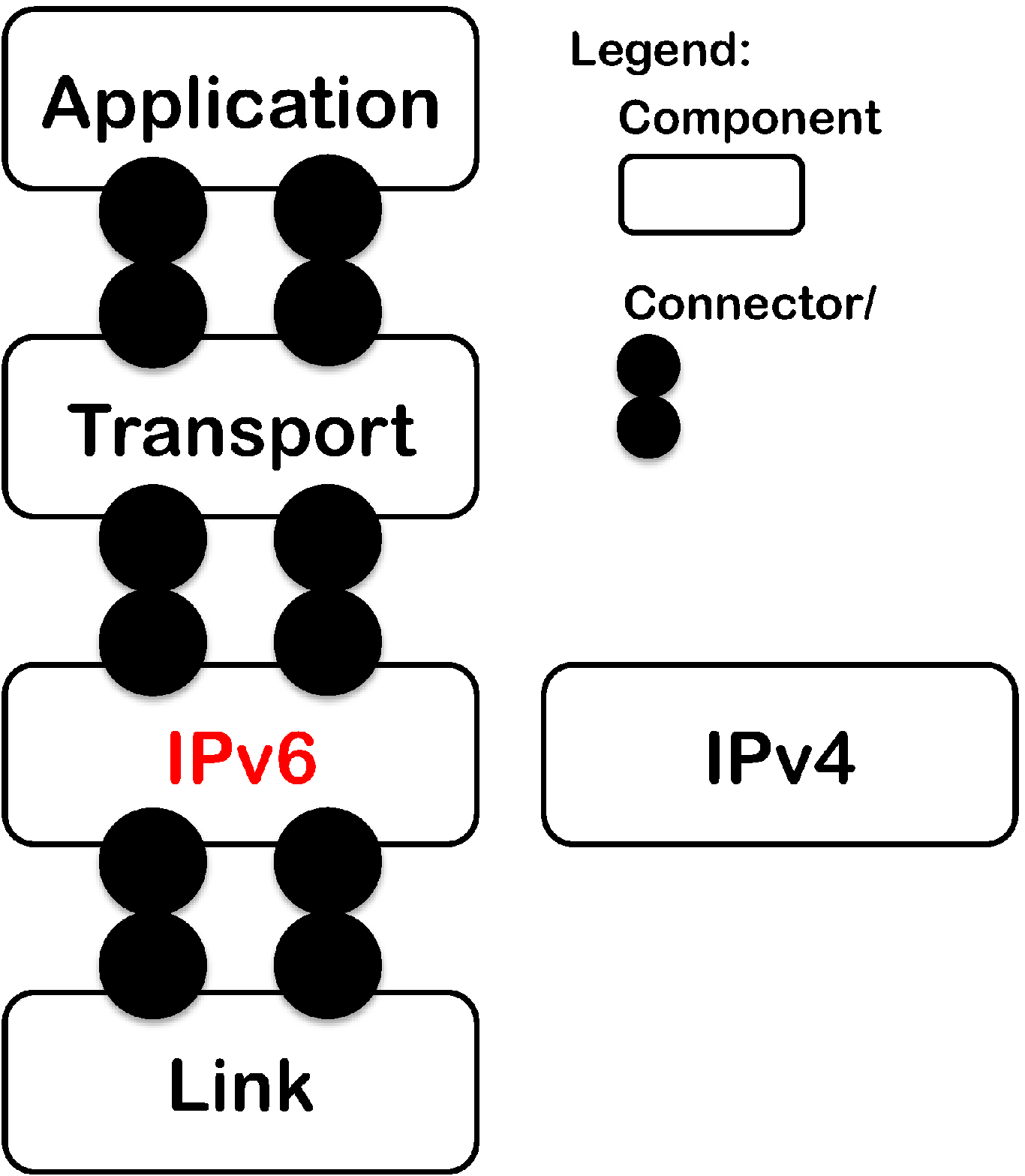, height=1.95in, width=1.70in}
   \label{fig:tcpip-ipv6}
}
\caption{scenario 2 to illustrate insertion of component.}\label{fig:tcpip-scenario2}
\end{figure}

We consider the three following reconfigurations:
\begin{enumerate}
   \item \emph{Switch the application component.}
      Assume two alternatives: MPEG-Decoder and H263-Decoder.
      MPEG-Decoder is used if \emph{bandwidth} is high (Fig.~\ref{fig:tcpip-mpeg}).
      Otherwise H263-Decoder is selected (Fig~\ref{fig:tcpip-h263}).
      Switching between MPEG and H263 starts a new stream.
      Therefore, no state in the decoder needs be kept. 
      We can therefore replace one component by another. 
   \item \emph{Insertion of a new component type.}
      Presume that the Internet Protocol version 6 (IPv6) replace the version 4 (IPv4) (Fig.~\ref{fig:tcpip-ipv6}).
      The IPv6 component has backward compatibility with the IPv4 component. 
      So it has the ports for both IPv4 (for compatibility reason) and for IPv6 (new feature). 
      Therefore, the type of the component at the internet layer is changed: it has 2 additional ports. 
   \item \emph{Update a component type.}
      Assume that the algorithm of error control of the Transport component is improved.
      When the error control algorithm is changed, the new behaviour needs to know the status of each packet (acknowledged, sent, timeout, ...) as well as the content of non-acknowledged packets (in order to resend them). 
      Therefore, only the behaviour part of the component should be replaced. 
      The state of the component should be unchanged.
\end{enumerate}

These scenarios cover a wide spectrum.
They can either be programmed at design-time or be discovered at run-time.
They feature reconfiguration at both type and instance level.
They target both the structure and the behaviour of the architecture.
We investigate the ADLs literature in order to illustrate how reconfiguration is supported in these scenarios.

In this paper, we focus on four ADLs that are representative of the main trends. 
Each ADLs complement the other.
The complement at the sense for describing an initial architecture as well as dynamic reconfiguration.
We choose $\pi$-ADL because it provides both description with behaviours thanks to its $\pi$-calculus roots.
Whilst the ACME/Plastik is a declarative language with focus on the architecture structure for both descriptions.
While C2SADL is used to compose architectures based on component and dynamic reconfiguration based on architectural events.
Finally, Dynamic Wright provides support for architectural description based in graph-grammars and dynamic reconfiguration specification with a variant of Communicating Sequential Processes (CSP).

\section{Underlying concepts}\label{section:concepts}

\paragraph{Software architecture concepts}\hfill Software architectures\\des\-cri\-be systems by specifying 
their elements and the interactions between 
them~\cite{Bass2003, Shaw1996}.
In this work, we use the basics concepts defined 
in~\cite{Garlan1997b, Medvidovic2000, Shaw1996}:
a \emph{component} is an unit of computation and storage; 
a \emph{connector} is an unit of communication; and, 
a \emph{configuration} is a specification of a software 
architecture in terms of components, connectors, and 
relationship between them.

Ports, roles, behaviours, constraints, and non-functional pro\-per\-ties are other concepts defined in~\cite{Garlan1997b, Medvidovic2000, Shaw1996} used to describe these architectural elements.
Ports and roles are used to define the way of interaction, ports for configurations and components, and roles for connectors.
Behaviours are the architectural element internal computation.
And the last two concepts, constraints and non-functional properties, are used to denote the assertions, invariants, quality of service that architectural elements are expected to meet.

Also, we consider types and instances of configurations, components, and connectors for modelling a software architectures~\cite{Medvidovic2000, Shaw1996}.
The types are abstractions like classes in the Object-Oriented Paradigm which encapsulate both the structure and the behaviours that can be performed on its structure.
These types will be used to build the instances and these instances will be used to describe software architectures.

Fig.~\ref{code:tcpip-acme} illustrates in ACME the basics concepts with our example TCP/IP stack system (Sec.~\ref{section:motivation} Fig.~\ref{fig:tcpip-system}) at instance and type level.
The types are defined in the \lstinline!Family! statement block (lines 1-18) while the instance level in \lstinline!System! (lines 20-30).

\begin{figure}[t]
\setlength{\fboxsep}{0pt}
\setlength{\fboxrule}{0pt}
\scriptsize\begin{lstlisting}[language=ACME,fontadjust,basewidth=.45em,xleftmargin=5.0ex]
Family TCPIP_MF {
  . . .
  Connector Type TCPIP_Conn2Layers {
    ProvidedRole source;
    RequiredRole sink;
  };
  Component Type TCPIP_Component {
    Port Type DataFrom extends ProvidedPort with {};
    Port Type DataTo extends RevidedPort with {};
    Property Type Layer = enum{application, transport, internet, link}
  };
  Component Type TCPIP_Application extends TCPIP_Component with {
    Port dataReceivedFromTransport: DataFrom;
    Port requiredService: DataTo;
    Property layer = "application";
  };
  . . .
}

System DecoderStream : TCPIP_MF = new TCPIP_MF extended with {
  Component application : TCPIP_Application;
  Component transport : TCPIP_Transpot;
  Component internet : TCPIP_Internet;
  Component link : TCPIP_Link;
   
  Connector 
    application2transport, transport2internet, internet2link: Conn2Layers;
  Attchments {
    application.requiredService to application2transport.source;
    application2transport.sink to transport.service;

    transport.requiredService to transport2internet.source;
    transport2internet.sink to internet.service;

    internet.requiredService to internet2link.source;
    internet2link.sink to link.service;
  }
  . . .
}
\end{lstlisting}\normalsize
\caption{Simplified version of TCP/IP stack system in ACME ADL.}
\label{code:tcpip-acme}
\end{figure}

\paragraph{Dynamic reconfiguration concepts}
A dynamic reconfiguration is a set of operations to modify an existing configuration at runtime.
At the software architecture level, the operation is defined in terms of the architectural elements at type or instance level.
Scenario 1 is an example of reconfiguration at the instance level (MPEG and H263 share the same type); and 
scenario 2 is an example of reconfiguration at the type level (IPv6 extends the type of IPv4).

Also, the dynamic reconfiguration can be foreseen or unforeseen~\cite{Gomes2007}.
The foreseen is a programmed dynamic reconfiguration specified at design time.
While the unforeseen is an ad-doc defined at runtime.
Example of foreseen and unforeseen is the scenario 1 and 3, respectively.
In the scenario 1, the architect specify at design time that the system modify itself when the bandwidth change.
Whereas the behaviour improvement is built at runtime as defined in scenario 3.

\section{Related Work}\label{section:related-works}
Medvidovic and Taylor~\cite{Medvidovic2000} created a classification framework based on architectural elements and illustrated them with the following ADLs: ACME, Aesop, C2, Darwin, Me\-taH, Rapide, SADL, UniCon, Weaves, and Wright.
However, this work has no focus on the dynamic reconfiguration.
Hence, its discussion about dynamic reconfigurations is limited, 
e.g. all ADLs were assessed only at the instance level, not at the type level.

Kacem \emph{et al.}~\cite{Kacem2002} described the capabilities of Darwin, ArchJava, Olan, Rapide, Wright, and ACME to support dynamic reconfiguration. 
They classify the ADLs as configuration and description language. 
The configuration language supports the description of a software system and limited dynamic reconfigurations, while the description language supports the specification of modifications to be applied to an existing architecture.
Despite this work focus is on dynamic reconfiguration, it also presents some limitations such as the fact that the evolution is applied only at instance level and do not take into account the behaviour of the architectural elements.

Bradbury~\cite{Bradbury2004} and Bradbury \emph{et al.}~\cite{Bradbury2004a} proposed  a framework to compare fourteen formal specification approaches which support dynamic software architectures. 
Amongst the criteria, Bradbury \emph{et al.} considered the basic operations, how they can be composed and whether they apply to variable sets of architectural element types.
About the basic operations (addition and removal of components and connectors), Bradbury et al. showed that most approaches support them.
Also, they considered the criterion of operations composition (basic\footnote{\emph{Basic composite operation} is the ability to group basic operations for reconfigurations.}, sequence, choice, and iteration).
Only 3 approaches (CommUnity, Dynamic Wright, and Gerel) provide full support while the other only support basic composition.
Finally, Bradury \emph{et al.} mentioned that all approaches support only the types of architectural elements defined prior runtime, fixed sets of architectural elements.

While ADL surveys have reached good understanding of the concepts underlying software architectures (Medvidovic and Taylor~\cite{Medvidovic2000}), this is not true regarding dynamic reconfiguration at the architecture level. 
Even Bradbury~\cite{Bradbury2004} and Bradbury \emph{et al.}~\cite{Bradbury2004a}, probably the most complete surveys, do not consider operations such as insertion of configuration and behaviour modification. 
They do not recognize \emph{link} and \emph{unlink} as basic operations.
However, they present the examples of these operations on some approaches, like Gerel and C2SADL.
Considering the scenarios of section~\ref{section:motivation}, these related works address only the first scenario (partially) but not the two last.
Consequently, it is clear that to discuss the three scenarios presented in section~\ref{section:motivation} is essential to have a better understanding of the dynamic reconfiguration issues.
These scenarios consider both instance and type levels, the basic operations include configurations, components, and connectors, as well as their structures and behaviours.

\section{Comparison of 4 ADL}\label{section:comparison}

\subsection{Applying the example in ADLs}
In this subsection, we describe the scenarios defined in Sec.~\ref{section:motivation} in all four ADLs: $\pi$-ADL (subsection~\ref{sec:pi-adl}), ACME/Plastik (subsection~\ref{sec:acme}), C2SADL (subsection~\ref{sec:c2sadl}), and Dynamic Wright (subsection~\ref{sec:wright}).
For each ADL we show the purpose, the TCP/IP stack system, a solution for three scenarios, and discussion about this implementation.

\subsubsection{$\pi$-ADL}\label{sec:pi-adl}
\paragraph{Purpose of the ADL} $\pi$-ADL is based on 
$\pi$-calculus and it is designed to describe the 
concurrent and mobile systems~\cite{Oquendo2004}.
The description of architectural elements is mainly 
represented by ports and behaviour.

\paragraph{Dynamic reconfiguration support}\hfill This ADL provides\\
support for foreseen and unforeseen dynamic reconfiguration.
The foreseen can be described in the behaviour of any 
architectural elements. While unforeseen needs some 
support, e.g., in the Virtual Machine (VM), in order 
to obtain a root reference to the reconfigured system.
In ArchWare, this issue is addressed by the deep intrication between the toolset (including the visual and textual editors) and the VM thanks to the hyper-code technology~\cite{Oquendo2004e}.

\paragraph{Initial architecture}
The implementation of TCP/IP stack system is shown in Fig.~\ref{code:tcpip-piadl}.
The configuration is composed at lines 1-18 basically with a behaviour, statement  \lstinline!behaviour! is ... where {...}! (lines 2-17).
This architectural behaviour defines the instances of components (lines 3-6) and connectors (lines 7-9).
The statement \lstinline!where! in behaviour specifies the connections between components and connectors (lines 11-16).
The example of component \lstinline!TCPIP_Application! is partially showed at lines 20-35: two ports and a behaviour.
The component behaviour is implemented with two internal methods (lines 24-25) and the definition of communication between its ports (lines 27-33).

\begin{figure}[t]
\setlength{\fboxsep}{0pt}
\setlength{\fboxrule}{0pt}
\scriptsize\begin{lstlisting}[language=PIADL,fontadjust,basewidth=.45em,xleftmargin=5.0ex]
archictecture TCPIP is abstraction() {
  behaviour is compose {
    application is TCPIP_Application()
    and transport  is TCPIP_Transport()
    and net        is TCPIP_Internet()
    and link       is TCPIP_Link()
    and app2transp is Conn2Layers()
    and transp2net is Conn2Layers()
    and net2link   is Conn2Layers()
  } where {
    appliation::request    unifies app2transp::source
    and app2transp::sink   unifies transport::service
    and transpost::request unifies transp2net::source
    and transp2net::sink   unifies net::service
    and net::request       unifies net2link::source
    and net2link::sink     unifies link::service
  }
}

component TCPIP_Application is abstraction() {
   port service is { ... }.
   port request is { ... }.
   behaviour is {
      processRequest is function(d: DataType):DataType {unobservable}.
      processResponse is function(d:DataType):DataType {unobservable}.

      choose {
         via service::wait receive entryData: DataType.
         via request::call send processRequest(entryData).
      or
         via request::response receive replyData: DataType.
         via service::reply send processResponse(replyData).
      }
   }
}

component TCPIP_Transport is abstraction() { ... }

component TCPIP_Internet is abstraction() { ... }

component TCPIP_Link is abstraction() { ... }

connector Conn2Layers is abstraction() {
  ...
  port source is { ... }
  port sink is { ... }

  behaviour is { ... }
}
\end{lstlisting}\normalsize
\caption{Partial specification of the TCP/IP stack system in $\pi$-ADL.}
\label{code:tcpip-piadl}
\end{figure}

\paragraph{Scenario 1}
The implementation of scenario 1 in $\pi$-ADL is composed by a component that performs the dynamic reconfiguration (Fig.~\ref{code:scenario1-piadl}).
This component has two parameters: the application component and the system component.
Both are represented as a behaviour.
The application component is a MPEG-Decoder or H263-Decoder.

Still, this component is structured with seven ``variables'' and a behaviour, lines 2-4 and 5-27, respectively.
The variables are used for behaviours of components and connectors.
To perform the modification, the behaviour component use the following steps:
\footnote{Decomposition, modification, and composition is a process provided by the ArchWare Virtual Machine. 
This facility is possible: 
take a snapshot of the system or a specific component/connector, both represented as a sequence of behaviour; 
make a modification; and, compose again the behaviour with new composition.
When it is composed, the ArchWare Virtual Machine automatically updates the system.}
decomposition the system behaviour (line 6);
assign the behaviours to variables (lines 8-14);
compose a new system behaviour (lines 16-26) with a replaced application component (line 20).

\begin{figure}[t]
\setlength{\fboxsep}{0pt}
\setlength{\fboxrule}{0pt}
\scriptsize\begin{lstlisting}[language=PIADL,fontadjust,basewidth=.45em,xleftmargin=5.0ex]
component reconfiguration is abstraction(application: Behaviour, system: Behaviour) {
   behaviours : sequence[Behaviour].
   transport : Behaviour. net : Behaviour. link : Behaviour.
   app2transp : Behaviour. transp2net : Behaviour. net2link : Behaviour.
   behaviour is {
      behaviours := decompose system.

      transport := behaviours::1::bhvr.
      net := behaviours::2::bhvr.
      link := behaviours::3::bhvr.

      app2transp := behaviours::4::bhvr.
      transp2net := behaviours::5::bhvr.
      net2link := behaviours::6::bhvr.

      compose {
         application and transport and net and link
         and app2transp and transp2net and net2link
      } where {
         appliation::request    unifies app2transp::source
         and app2transp::sink   unifies transport::service
         and transpost::request unifies transp2net::source
         and transp2net::sink   unifies net::service
         and net::request       unifies net2link::source
         and net2link::sink     unifies link::service
      }
   }
}
\end{lstlisting}\normalsize
\caption{Implementation of scenario 1 in $\pi$-ADL.}
\label{code:scenario1-piadl}
\end{figure}

\paragraph{Scenario 2}
The implementation for the unforeseen scenario 2 is shown in Fig.~\ref{code:scenario2-piadl}.
This code is applied in a system at runtime to deploy the two new component ``type''.
First, the component type is a new protocol \emph{IPv6} (lines 1-3), and the second is a component to perform dynamic reconfiguration (lines 4-32).
In order to execute this dynamic reconfiguration two steps are applied: (i) to obtain the root behaviour of the system, and (ii) to invoke the computation of the reconfiguration component.
Both operations are aided by toolset and hyper-code.

The code of the reconfiguration component is described using the following steps: 
(i) to decompose the behaviours of the system in components and connectors (lines 9); (ii) to
assign to ``variables'' the instances of components that represent application, transport, and link (lines 11-13); (iii) to
attribute connectors to ``variables'' (lines 15-17); (iv) to
compose a new system behaviour (lines 19-30) with a new instance of the protocol IPv6 component (line 21).
As in the scenario 1, when performing a new composition, the behaviour of the system is automatically updated.

\begin{figure}
\setlength{\fboxsep}{0pt}
\setlength{\fboxrule}{0pt}
\scriptsize\begin{lstlisting}[language=PIADL,fontadjust,basewidth=.45em,xleftmargin=5.0ex]
component TCPIP_IPv6 is abstracion() {
  ...
}
component reconfiguration is abstraction(system: Behaviour) {
   behaviours : sequence[Behaviour].
   application: Behaviour. transport : Behaviour. link : Behaviour.
   app2transp : Behaviour. transp2net : Behaviour. net2link : Behaviour.
   behaviour is {
      behaviours := decompose system.

      application := behaviours::0::bhvr.
      transport := behaviours::1::bhvr.
      link := behaviours::3::bhvr.

      app2transp := behaviours::4::bhvr.
      transp2net := behaviours::5::bhvr.
      net2link := behaviours::6::bhvr.

      compose {
         application and transport and link
         and net is TCPIP_v6
         and app2transp and transp2net and net2link
      } where {
         appliation::request    unifies app2transp::source
         and app2transp::sink   unifies transport::service
         and transpost::request unifies transp2net::source
         and transp2net::sink   unifies net::service
         and net::request       unifies net2link::source
         and net2link::sink     unifies link::service
      }
   }
}
\end{lstlisting}\normalsize
\caption{Implementation of scenario 2 in $\pi$-ADL.}
\label{code:scenario2-piadl}
\end{figure}

\paragraph{Scenario 3}
In the scenario 3 it is also used the process of decomposition, modification, and composition, as showed above.
We use the ArchWare Virtual Machine facility to get a reference to the root of the target Transport component.
After, we make a modification into the behaviour of the component.
At last, the ArchWare Virtual Machine synchronize the system with this modification.

%

\paragraph{Summary of $\pi$-ADL}
With these implementations, some issues are identified on $\pi$-ADL support for dynamic reconfiguration:
\begin{enumerate}
	\item changing types needs external help, e.g. toolset, hyper-code and $\pi$-ARL. Toolset and hyper-code are used for capturing, specifying the changes, and applying the changes. While $\pi$-ARL is a language to specify refinements in the system;
	\item instances modifications can be specified in architectural elements or also with toolset and hyper-code help. At first, usually the dynamic reconfiguration is tangled with nominally behaviour. However, we used other approach with a specific component to specify dynamic reconfiguration;
	\item For the third scenario, the $\pi$-ADL do not provide a mechanism to design the intermediate states for updating the instances with the improvement of behaviour. However, if we use the $\pi$-ARL approach, we can have better control;
	\item Also with $\pi$-ARL is possible build constraints and non-functional properties for architectural elements. However, it's not used to assess dynamic reconfiguration.
\end{enumerate}

\subsubsection{ACME ADL and Plastik}\label{sec:acme}
\paragraph{Purpose of the ADL}
ACME/Armani is a declarative language based on the first-order predicate logic~\cite{Garlan2000, Monroe2001}.
Its initial purpose was to define a common interchange language for architecture design tools.
Also, its statements are designed to describe the architectural structures at instance and type levels.
The extensions~\cite{Batista2008a, Joolia2005} to support dynamic reconfiguration preserves these initials purpose, and are named as ACME/Plastik.
Thereafter, any configuration and dynamic reconfiguration specified using this extension (ACME/Plastik) are based on the structure of an architecture.

\paragraph{Dynamic reconfiguration support}
ACME/Plastik permits to describe foreseen and unforeseen dynamic reconfiguration.
To unforeseen reconfiguration it is possible to describe as a specific behaviour of a configuration with the \lstinline!on ($conditional\_expression$) do {$operations$}! statement.
Unforeseen reconfiguration can be expressed in a script that will be applied in the system at runtime aided by a toolset provided by the Plastik framework.
Both, $operations$ and script are composed with ACME statements and its ACME/Plastik extension.
The $conditional\_expression$ is composed in the Armani language~\cite{Monroe2001}.

\paragraph{Initial architecture}
The example of the TCP/IP stack system in ACME is showed in Fig.~\ref{code:tcpip-acme}.
The types are specified in the \lstinline!Family! statement while the instances are in the \lstinline!System! statement.
For types, a connector (lines 3-6) and two components (lines 7-11 and 12-16) are implemented as example.
The configuration is composed by the component instances (lines 21-24), connector instances (lines 26-27), and the connections between them (lines 28-37).

\paragraph{Scenario 1}
The implementation in ACME/Plastik of scenario 1 is showed in Fig.~\ref{code:scenario1-acme}.
The configuration is named with \lstinline!StreamDecoderSystem! and it extends the \lstinline!TCPIP_MF! and \lstinline!PlastikMF!.
The \lstinline!TCPIP_MF! is showed in Fig.~\ref{code:tcpip-acme} while \lstinline!Plastik_MF! in~\cite{Batista2008a, Joolia2005}.
Such configuration consists of: the
insertion of two new components (type and instance) in lines 2-5 and line 6;
the instantiation of components and connectors (omitted with comment and $...$ in lines 7-8);
the linking of components and connectors (omitted with comment and $...$ in lines 7-8);
and the specification of two dynamic reconfiguration situations, lines 12-19 and 21-28.

Both dynamic reconfigurations use the same similar description.
First, the conditional expressions are \lstinline!link.bandwidth == 'low'! and \lstinline!link.bandwidth == 'high'!.
Second, the $operations$ are described as two unlinking (lines 13-14, 22-23) and one linking (lines 15-18, 24-27) statements.
These $operations$ specified the replacement of an instance of the component.

\begin{figure}
\setlength{\fboxsep}{0pt}
\setlength{\fboxrule}{0pt}
\scriptsize\begin{lstlisting}[language=ACME,fontadjust,basewidth=.45em,xleftmargin=5.0ex]
System StreamDecoderSystem : TCPIP_MF, PlastikMF {
  Component mpeg-decoder : TCPIP_MpegDecoder = new TCPIP_Application extend with {
    property decoder-type = "MPEG";
    property algorithm = ...;
  }
  Component h263-decoder : TCPIP_H263Decoder = new TCPIP_Application extend with { ... }
// other instances of component and connector
  ...
// initial attachments
  ...
// programmed dynamic reconfiguration to low bandwidth
  on (link.bandwidth == 'low') do {
    detach mpeg-decoder.requiredService to application2transport.source;
    detach transport2application.sink to mpeg-decoder.dataReceivedFromTransport;
    attachments {
      h263-decoder.requiredService to application2transport.source;
      transport2application.sink to h263-decoder.dataReceivedFromTransport;
    }
  }
// programmed dynamic reconfiguration to high bandwidth
  on (link.bandwidth == 'high') do {
    detach h263-decoder.dataTo to application2transport.source;
    detach transport2application.sink to h263-decoder.dataReceivedFromTransport;
    attachments {
      mpeg-decoder.requiredService to application2transport.source;
      transport2application.sink to mpeg-decoder.dataReceivedFromTransport;
    }
  }
}
\end{lstlisting}\normalsize
\caption{Implementation of scenario 1 in ACME/Plastik.}
\label{code:scenario1-acme}
\end{figure}

\paragraph{Scenario 2}
The scenario 2, an unforeseen insertion of a new component type, is describe with an ACME/Plastik script in Fig.~\ref{code:scenario2-acme}.
For this scenario, the script is composed by: the insertion of a type and instance of a component (lines 1-3), the unlinking of the old component instance (lines 5-6) and the linking of the new component (lines 7-10).

\begin{figure}
\setlength{\fboxsep}{0pt}
\setlength{\fboxrule}{0pt}
\scriptsize\begin{lstlisting}[fontadjust,basewidth=.45em,xleftmargin=5.0ex]
Component ipv6 : TCPIP_IPv6 = new TCPIP_Internet extended with {
  ...
}

detach ipv4.requiredService to internet2link.source;
detach link2internet.sink to ipv4.dataReceivedFromLink;
attachments {
  ipv6.requiredService to internet2link.source;
  link2internet.sink to ipv6.dataReceivedFromLink;
}
\end{lstlisting}\normalsize
\caption{Implementation of scenario 2 in ACME/Plastik}
\label{code:scenario2-acme}
\end{figure}

\paragraph{Scenario 3}
The unforeseen upgrade of a behaviour in scenario 3 can be described by the script illustrated in Fig.~\ref{code:scenario3-acme}.
This script defines a new version of a component with an improvement on its behaviour.
The behaviour in ACME/Plastik is expressed as a property and it can be also described in other languages, such as external formal languages, see lines 3-5 in Fig.~\ref{code:scenario3-acme}.

\begin{figure}
\setlength{\fboxsep}{0pt}
\setlength{\fboxrule}{0pt}
\scriptsize\begin{lstlisting}[fontadjust,basewidth=.45em,xleftmargin=5.0ex]
Component Type TCPIP_Transport extends TCPIP_Component with {
  ...
  Property behaviour = { 
    \\ description of behaviour in other languages, e.g. CSP
  }
  ...
}
\end{lstlisting}\normalsize
\caption{Implementation of scenario 3 in ACME/Plastik}
\label{code:scenario3-acme}
\end{figure}

\paragraph{Summary of ACME/Plastik}
The dynamic reconfiguration issues for ACME can be summarized as follows:
\begin{enumerate}
	\item Dynamic reconfiguration in terms of structure is supported by ACME/Plastik. In terms of behaviour, dynamic reconfiguration is described using any external language embedded inside the property element. Usually, this is used for foreseen reconfiguration. 
This approach is limited in the sense that components and connectors usually needs the dynamic reconfiguration of their behaviour but as ACME/Plastik does not provides elements for behaviour specification, the architect has to rely on an external language. 
E.g. if the \lstinline!TCPIP_Transport! component needs to decide if it has a buffer or not, and if it uses a error detecting or not, and so on. These situations have to be described using an external language (in general formal languages such as CSP).

	\item ACME/Plastik does not provide a mechanism to control the intermediate states of a reconfiguration. 
      E.g. the implementation of the scenario 3, the architect cannot define the approach used to update the instances.
	\item For unforeseen reconfiguration ACME/Plastik relies on external scripts. This approach requires an external interpreted associated to the ADL to interpret the external script in order to reconfigure the system. 
\end{enumerate}

\subsubsection{C2SADL}\label{sec:c2sadl}
\paragraph{Purpose of the ADL}
Initially, the purpose of C2 was to describe the architecture of a software based on a Graphical User Interface (GUI)~\cite{Taylor1996a}.
Therefore, the ADL is based on hierarchically concurrent components.
C2 also allows the use of messages to notify the architectural elements.
C2 is subdivided in 2 languages: C2 IDL (Interface Description Language) for describing the components types, and C2 ADL to specify the configurations.
For components it is possible to specify top and bottom ports, to declare internal methods, and to specify the behaviour of its ports.
The implementation of the internal methods is done by the developer in the source code generated by a toolset.
With C2 ADL it is possible to define instances of components and connectors, as well as links between them.

\paragraph{Dynamic reconfiguration support}\hfill The Architectural\\Modification Language (AML) was created to extend C2 for supporting dynamic reconfiguration~\cite{Medvidovic1996, Oreizy1998}.
AML is a declarative language that uses the structure view of an existing configuration and messages to notify this configuration about a dynamic reconfiguration.
This language provides statements to build the scripts that specify dynamic reconfiguration.
However, in these scripts it is not possible to determine the moment to apply a dynamic reconfiguration.
Because that, human intervention is needed via a toolset to trigger dynamic reconfiguration.

\paragraph{Initial architecture}
The TCP/IP stack system is descri\-bed in C2 IDL in Fig.~\ref{code:tcpip-c2idl} and C2 ADL in Fig.~\ref{code:tcpip-c2adl}.

\begin{figure}
\setlength{\fboxsep}{0pt}
\setlength{\fboxrule}{0pt}
\scriptsize\begin{lstlisting}[language=C2SADL,fontadjust,basewidth=.45em,xleftmargin=5.0ex]
   ...
   component TCPIP_Transport is
      interface // define the component ports
         top_domain is // ports to link high level connectors
            out
               ...
            in
               ReceiveData(package: ApplicationPackage);
         bottom_domain is // ports to link low level connectors
            out
               SendToInternt(package: TransportPackage);
            in
               ...
      methods // define interfaces of internal behaviours
         function pack(data: ApplicationPackage) : TransportPackage);
         ...
      behaviour // describe the external behaviour 
         ...
         received_messages ReceiveData;
            invoke_methods Pack;
            always_genarate SendToInternet;
         ...
   end TCPIP_Transport
   ...
\end{lstlisting}\normalsize
\caption{Components of TCP-IP stack system definition in C2 IDL.}
\label{code:tcpip-c2idl}
\end{figure}

\begin{figure}
\setlength{\fboxsep}{0pt}
\setlength{\fboxrule}{0pt}
\scriptsize\begin{lstlisting}[language=C2SADL,fontadjust,basewidth=.45em,xleftmargin=5.0ex]
   architecture DecoderStream is
      component_intances {
         mpeg instantiates TCPIP_MpegDecoder;
         h263 instantiates TCPIP_H263Decoder;
         transport instantiates TCPIP_Transport;
         internet instantiates TCPIP_Internet;
         link instantiates TCPIP_Link;
      }
      connectors {
         application2transport;
         transport2internet;
         internet2link;
      }
      topology {
         connector application2transport {
            top_ports { mpeg filter no_filtering; }
            bottom_ports { transport filter no_filtering; }
         }
         connector transport2internet {
            top_ports { transport filter no_filtering; }
            bottom_ports { internet filter no_filtering; }
         }
         connector internet2link {
            top_ports { internet filter no_filtering; }
            bottom_ports { link filter no_filtering; }
         }
      }
   end DecoderStream;
\end{lstlisting}\normalsize
\caption{TCP-IP stack system implementation in C2 ADL.}
\label{code:tcpip-c2adl}
\end{figure}

\paragraph{Scenario 1}
For scenario 1, for changing the component according to the low or high bandwidth state, it is necessary two C2SADL scripts.
In Fig.~\ref{code:scenario1-c2sadl} is described the script for dynamic reconfiguration when bandwidth is low.
This script uses two statements, a \lstinline!Unweld! for unlinking the $mpeg$ instance and \lstinline!Weld! for linking the $h263$ instance.
Other AML's statements are \lstinline!AddComponent!, \lstinline!AddConnector!, \lstinline!RemoveComponent!, and \lstinline!RemoveConnector!.
All these statements are used with a defined configuration, e.g. line 2 and 3 in Fig.~\ref{code:scenario1-c2sadl} with the configuration named as \lstinline!DecoderStream!.
Despite of the fact that AML statements are based on services of the ArchStudio tool suite,
there are services that are not provided by the AML language, e.g.  \lstinline!start()! and \lstinline!stop()!.

\begin{figure}
\setlength{\fboxsep}{0pt}
\setlength{\fboxrule}{0pt}
\scriptsize\begin{lstlisting}[language=C2SADL,fontadjust,basewidth=.45em,xleftmargin=5.0ex]
\\ Bandwidth is low
DecoderStream.Unweld(mpeg, application2transport);
DecoderStream.Weld(h263, application2transport);
\end{lstlisting}\normalsize
\caption{Partial implementation of scenario 1 in C2SADL.}
\label{code:scenario1-c2sadl}
\end{figure}

\paragraph{Scenario 2}
The second scenario, the unforeseen insertion of component instance and type, is inferred as shown in Fig.~\ref{code:scenario2-c2sadl}.
Inference because both work about AML~\cite{Medvidovic1996, Oreizy1998} stated that it is possible to make insertion and deletion of types.
However, they only show examples how to build dynamic reconfiguration at the instance level.
Medvidovic \emph{et al.}~\cite{Medvidovic1996b} mention types and subtypes but at design level.

Dynamic reconfiguration in Fig.~\ref{code:scenario2-c2sadl} specifies:  the insertion of a new type of component (lines 2-5), the creation a new instance of component (line 9), the unlinking of the instance of IPv4 (lines 12-13), and the linking of the instance of IPv6 (lines 16-17).
The component type name is inferred by a toolset before performing dynamic reconfiguration.

\begin{figure}
\setlength{\fboxsep}{0pt}
\setlength{\fboxrule}{0pt}
\scriptsize\begin{lstlisting}[language=C2SADL,fontadjust,basewidth=.45em,xleftmargin=5.0ex]
\\ Creata a new type of component
component TCPIP_IPv6 is subtype
  all <= all TCPIP_Internet (...)
  ...
end TCPIP_IPv6

// Create a new instance of component
//   The name of the type is implicitly discovered
DecoderStream.AddComponent(TCPIP_IPv61);

// Unlink the instance of component IPv4
DecoderStream.Unweld(trasport2internet, TCPIP_IPv41);
DecoderStream.Unweld(TCPIP_IPv41, internet2link);

// Link the instance of component IPv6
DecoderStream.Weld(trasport2internet, TCPIP_IPv61);
DecoderStream.Weld(TCPIP_IPv61, internet2link);
\end{lstlisting}\normalsize
\caption{Implementation of scenario 2 in C2SADL.}
\label{code:scenario2-c2sadl}
\end{figure}

\paragraph{Scenario 3}
The third scenario is an improvement of the behaviour that can be modeled by the \lstinline!behaviour! statement via a sequence of invocations to internal methods.
Example of this change, see line 6 in Fig.~\ref{code:scenario3-c2sadl}, the invocation is modified to verify the received and packed data before sending a transport package to the internet layer.
If the change is an improvement of an internal method, it is not possible to describe with C2SADL.

\begin{figure}
\setlength{\fboxsep}{0pt}
\setlength{\fboxrule}{0pt}
\scriptsize\begin{lstlisting}[language=C2SADL,fontadjust,basewidth=.45em,xleftmargin=5.0ex]
   component TCPIP_Transport is
      ...
      behaviour // describe the external behaviour 
         ...
         received_messages ReceiveData;
            invoke_methods VerifyData, Pack;
            always_genarate SendToInternet;
         ...
   end TCPIP_Transport
   ...
\end{lstlisting}\normalsize
\caption{Implementation of scenario 3 in C2SADL.}
\label{code:scenario3-c2sadl}
\end{figure}

\paragraph{Summary of C2SADL}
After the C2SADL implementation of the three scenarios, we identified these issues:
\begin{enumerate}
	\item ArchStudio tool provides some services for supporting dynamic reconfiguration, 
      e.g. to start and to stop architectural elements. However, these services are not available in AML. Thus, the dynamic reconfiguration specification in AML is limited.  

   \item foreseen dynamic reconfiguration needs human intervention
      because C2SADL do not provide statements for the system apply dynamic reconfiguration.
	\item it does not support the specification of dynamic reconfiguration inside of the architectural elements. 
      Because that, the foreseen dynamic reconfiguration is implemented only at the implementation level;
	\item it does not provide a mechanism for controlling and monitoring the intermediate states of a dynamic reconfiguration;
	\item it also does not support the assessment of dynamic reconfiguration.
\end{enumerate}

\subsubsection{Dynamic Wright}\label{sec:wright}
\paragraph{Purpose of the ADL}
Dynamic Wright has focus on the structure and behaviour of an architecture~\cite{Allen1998a}.
Dynamic Wright supports the description of the architectural elements types structures and behaviours\footnote{In Dynamic Wright it is possible to specify the behaviour for component ports and computations, and connector roles and glues.}, and initial configuration.
Structure, initial configuration, instances, and links are described in a declarative form, while behaviours are specified in a graph-based and a variant of the Communicating Sequential Process (CSP) form.

\paragraph{Dynamic reconfiguration support}
Dynamic Wright only supports foreseen dynamic reconfiguration.
It is built as a special behaviour of the initial configuration.
It has the same base of a ``nominally'' behaviour of architectural elements.
The extension described in ~\cite{Allen1998a} proposes additional statements and control events.
The statements are \lstinline!new!, \lstinline!remove!, \lstinline!attach!, and \lstinline!detach!.
The control events are used to define the specific moment to perform a dynamic reconfiguration.

As Dynamic Wright does not support unforeseen reconfiguration, the scenarios 2 and 3 cannot be implemented.

\paragraph{Initial architecture}
The initial configuration of the TCP/\-IP stack system is illustrated in Fig.~\ref{code:scenario1-wright}.
The component and connector types are specified in the \lstinline!Style ... EndStyle! (lines 1-16) statements.
The implementation of the \lstinline!TCPIP_!\lstinline!MPEG-Decoder! Component is described with two ports (lines 3-4) and a behaviour (lines 6-7).
The configuration is described with the \lstinline!Configuror ... EndConfiguror! statement.
The initial configuration instances and connections are specified at lines 18-27.

\paragraph{Scenario 1}
Fig.~\ref{code:scenario1-wright} also illustrates the specification of scenario 1.
It uses the \lstinline[language=Wright]!Configuror [initial configuration] where [behaviours]! Statement.
\emph{behaviours} are special named behaviours of the configuration for specifying dynamic reconfiguration.
In this case, there are two specification: lines 29-36 for low bandwidth and lines 37-44 for high bandwidth.
The following steps were used to define the first and the second dynamic reconfiguration situations:

\begin{enumerate}
	\item to define a name (lines 29 and 37).
	\item to specify the control events that define the moment for performing the operations of dynamic reconfiguration (lines 30 and 38).
	\item to determine the type of configuration to use (lines 31 and 39). Dynamic Wright use the term \lstinline!Style! for this purpose.
	\item to specify the operations to perform dynamic reconfiguration (lines 32-35 and 40-43).
	\item to use the statements of the external choice ($\Box$) and successfully terminate ($\S$), lines 36 and 44.
\end{enumerate}

\begin{figure}[t]
\setlength{\fboxsep}{0pt}
\setlength{\fboxrule}{0pt}
\scriptsize\begin{lstlisting}[language=Wright,fontadjust,basewidth=.45em,xleftmargin=5.0ex]
Style TCPIP-Style
   Component TCPIP_MPEG-Decoder
      Port dataReceivedFromTransport = ...
      Port requiredService = ... 
         [describe the port behaviour using a variant of CSP]
      Computation = ... 
         [describe the component behaviour using a variant of CSP]
   ...
   Connector Conn2Layers
      Role source = ...
      Role sink = ...
      Glue = ... 
         [describe the connector behaviour using a variant of CSP]
   Constraints
      ...
EndStyle
Configuror DecoderStream
  Style TCPIP-Style
    new.mpeg : TCPIP_MPEG-Decoder
    $\rightarrow$ new.h263 : TCPIP_H263-Decoder
    ... [other component instances]
    $\rightarrow$ new.app2trans : Conn2Layers
    $\rightarrow$ new.trans2net : Conn2Layers
    $\rightarrow$ new.net2link : Conn2Layers
    $\rightarrow$ attach.mpeg.requiredService.to.app2trans.source
    $\rightarrow$ attach.app2trans.sink.to.transport.service
    ... [other attachments]
  where
    WaitForBandwidthLow = (
    link.control.bandwidthDown $\rightarrow$ mpeg.control.off $\rightarrow$ h263.control.on
    $\rightarrow$ Style TCPIP-Style
      detach.mpeg.requiredService.to.app2trans.source
      $\rightarrow$ attach.h263.requiredService.to.app2trans.source
      $\rightarrow$ detach.trans2app.sink.to.mpeg.dataReceivedFromTransport
      $\rightarrow$ attach.trans2app.sink.to.h263.dataReceivedFromTransport
    ) $\Box$ $\S$
    WaitForBandwidthHigh = (
       link.control.bandwidthUp $\rightarrow$ h263.control.off $\rightarrow$ -> mpeg.control.on
       $\rightarrow$ Style TCPIP-Style
          detach.h263.requiredService.to.app2trans.source
          $\rightarrow$ attach.mpeg.requiredService.to.app2trans.source
          $\rightarrow$ detach.trans2app.sink.to.h263.dataReceivedFromTransport
          $\rightarrow$ attach.trans2app.sink.to.mpeg.dataReceivedFromTransport
     ) $\Box$ $\S$
EndConfiguror
\end{lstlisting}\normalsize
\caption{Implementation of scenario 1 in Dynamic Wright}
\label{code:scenario1-wright}
\end{figure}

\paragraph{Summary of Dynamic Wright}
The following issues are identified after the implementation of the scenario 1:
\begin{enumerate}
	\item it does not provide operations to define dynamic reconfiguration for the type level of architectural elements;
	\item this ADL provides a refined mechanism of control e\-vents to determine the moment to perform a dynamic reconfiguration. Although, it does not provide a mechanism to control and monitor the intermediate states. e.g. if the application of the scenario 1 is a banking system, the state must be copied from the instance of the replaced component to the new component. Dynamic Wright 
does not provide operations for this purpose. 
	\item despite of providing the specification of behaviour for components and connectors, it is not possible to define dynamic reconfiguration for this architectural elements;
	\item similarly to the other ADLs, Dynamic Wright does not provide a mechanism for the assessment of dynamic reconfiguration.
\end{enumerate}

\subsection{Summary of comparison}
The summary of the four ADLs support for dynamic reconfigurations is shown in Tab.~\ref{tab:reconfiguration}.
$\pi$-ADL and ACME/Plastik are the only ADLs that support both foreseen and unforeseen dynamic reconfigurations.
As C2SADL with the AML language do not provide a way to specify an internal initiation of dynamic reconfiguration, 
foreseen dynamic reconfiguration cannot be automatically triggered.
Dynamic Wright supports only foreseen reconfiguration by defining the special behaviour of a configuration.

\begin{table}
\centering
\caption{Foreseen and unforeseen dynamic reconfiguration support in ADLs.}
\label{tab:reconfiguration}
\begin{tabular}{|p{0.5in} |p{1.25in} |p{1.25in} |}
\cline{2-3}
	\multicolumn{1}{l|}{} & Foreseen & Unforeseen \\
\hline
	$\pi$-ADL & 
	inside of all architectural elements behaviour & 
	ADL statements aided by the ArchWare toolset \\
\hline
	ACME Plastik &
	special behaviours in the configuration &
	ADL statements and external language aided by the Plastik framework \\
\hline
	C2SADL & 
	limited support by the ADL & 
	external script in AML \\
\hline
	Dynamic Wright & 
	special behaviours in the configuration & 
	 not provided by the ADL \\
\hline
\end{tabular}
\end{table}

The four ADLs do not address the issue of consistent reconfiguration (reconfiguration at type and instance levels) as shown the Tab.~\ref{tab:level}.
$\pi$-ADL and ACME provides operations for both levels, however the two ADLs use external help to specify reconfiguration at the type level. C2SADL AML provides operations to modify instances. In spite of the toolset provides the services for both levels.
Dynamic Wright does not provide operations to specify dynamic reconfiguration for the type of architectural elements.

\begin{table}
\centering
\caption{Dynamic reconfiguration support for modification on type and instance level.}
\label{tab:level}
\begin{tabular}{|p{0.5in} |p{1.25in} |p{1.25in} |}
\cline{2-3}
	\multicolumn{1}{l|}{} & Type & Instance \\
\hline
	$\pi$-ADL & 
	ADL statements with external help & 
	ADL statements \\
\hline
	ACME Plastik &
	ACME/Plastik statements with external help &
	ACME/Plastik statements \\
\hline
	C2SADL & 
	not provided by the ADL & 
	operations defined in AML \\
\hline
	Dynamic Wright & 
	not provided by the ADL & 
	operations defined in an extension of the ADL \\
\hline
\end{tabular}
\end{table}

The support for specifying the behaviour of the architectural elements  can be subdivided in two categories: nominally and dynamic reconfiguration.
Nominally is a set of functions that the architectural elements can  invoke. 
Dynamic reconfiguration is the specification of a behaviour to compose a foreseen dynamic reconfiguration.
The ADLs support for both is resumed in Tab.~\ref{tab:behaviour}.
$\pi$-ADL supports both in the architectural element behaviour.
ACME/Plastik nominally behaviour can be considered in two ways: a property of architectural element described with an external language, e.g. CSP; and, port/rule communication for components and connectors; for dynamic reconfiguration, ACME/Plastik provides the on-do clause. 
Dynamic Wright glue and computation clauses define the nominally behaviour of connectors and components; a configuration has a special behaviour for dynamic reconfiguration.
C2SADL has no behaviour for dynamic reconfiguration but nominally component behaviour is described by events communication.

\begin{table}
\centering
\caption{ADL support for behaviour specification.}
\label{tab:behaviour}
\begin{tabular}{|p{0.5in} |p{1.25in} |p{1.25in} |}
\cline{2-3}
	\multicolumn{1}{l|}{} & Nominally & Dynamic reconfiguration \\
\hline
	$\pi$-ADL & 
	All architectural elements with $\pi$-ADL behaviour. & 
	tangled with nominally behaviour. \\
\hline
	ACME Plastik &
	External language used in properties of the architectural elements. &
	ACME/Plastik on-do statement. \\
\hline
	C2SADL & 
	Component behaviour based on event communication.  & 
	Not provided. \\
\hline
	Dynamic Wright & 
	Variant of CSP in connector glue and component computation. & 
	Configuration special behaviour with variant of CSP. \\
\hline
\end{tabular}
\end{table}

None of the four ADLs has a mechanism for assessing the system.
$\pi$-ADL can rely on $\pi$-AAL for defining and analyzing the architectural constraints and non-functional properties.
While ACME/Plastik has Armani. even though it has no formal semantics, and therefore, it is not liable to rigorous analysis~\cite{Waewsawangwong2004}.
C2SADL provides type verification for events communication.
Dynamic Wright provides a refined mechanism of control events to determine the moment to perform a dynamic reconfiguration.
Although, the assessment of dynamic reconfiguration is not provided on four ADLs.

\section{Conclusion}\label{section:conclusion}
In this paper we analysed the support of ADLs for handling dynamic reconfigurations.
We started with a motivation example and some reconfiguration scenarios and we described the example and the scenarios using four well-known ADLS. 
We used the example and the specifications to discuss the following issues: 
(i) how each ADL addresses the issue of consistently reconfiguring both instances and types; 
(ii) how each ADL takes into account the behaviour of the architectural elements during reconfiguration; and 
(iii) how each ADL supports the assessing of dynamic reconfiguration.

In comparison to related works, we analysed some important issues for the dynamic reconfiguration support at the architectural level: 
foreseen and unforeseen changes; 
instance and type level modifications; 
structure and behaviour for all architectural elements; 
definition of nominally and dynamic reconfiguration behaviour for all architectural elements; and, 
the analysis of dynamic reconfiguration.

We can conclude that some issues still remain open as no ADL provides support for all of them together: how to apply the changes made in type level to their respective instances?
how to control the set of intermediate states of a software system during a dynamic reconfiguration?
how to assess dynamic reconfiguration?  

\section{Acknowledgments}
Special thanks to Coll\`ege Doctoral International (CDI)/Uni-versit\'e Europ\'eenne de Bretagne  (UEB) for financial support this work.

\bibliographystyle{abbrv}

\end{document}